\documentclass[final,1p,times]{elsarticle}
\usepackage{amssymb}

\journal{European Journal of Mechanics - B/Fluids}

\begin{document}

\begin{frontmatter}

%\title{Title\tnoteref{label1}}
%\tnotetext[label1]{}
%\author{Name\corref{cor1}\fnref{label2}}
%\ead{email address}
%\fntext[label2]{}
%\cortext[cor1]{}
%\address{Address\fnref{label3}}
%\fntext[label3]{}

\title{On the evolution of particle-puffs in
    turbulence\footnote{Available online 28 August 2015 doi:10.1016/j.euromechflu.2015.06.009}}

\author[label1]{S. Bianchi}
\author[label2]{L. Biferale}
\author[label3]{A. Celani}
\author[label4]{M. Cencini}
\address[label1]{Dept. Physics, University of Rome 'Roma Tre', via della Vasca Navale 84, 00146 Roma, Italy}
\address[label2]{Dept. Physics and INFN, University of Rome 'Tor Vergata', via della Ricerca Scientifica 1, 00133 Roma, Italy}
\address[label3]{The Abdus Salam International Centre for Theoretical Physics (ICTP), Strada Costiera 11, I-34014 Trieste, Italy}
\address[label4]{Istituto dei Sistemi Complessi, CNR, via dei Taurini 19, 00185 Roma, Italy}

%\author{}

%\address{}

\begin{abstract}

We study the evolution of turbulent puffs by means of high-resolution numerical simulations.
Puffs are bunches of passive particles released from an initially spherical distribution at 
regular time intervals of the order of the Kolmogorov time. The instantaneous shapes of particle puffs, in particular their asphericity and prolateness, are characterized by measuring the gyration tensor.  Analysis has been performed by
following, up to one large scale
eddy-turn-over time, more than $10^4$ different puffs, each made of 2000 particle tracers, emitted from different places in a homogeneous and
isotropic turbulent fluid with Taylor-scale Reynolds number $Re_\lambda \sim
300$. We also analyze the probability of hitting a given target placed
downstream with respect to the local wind at the time of emission,
presenting data for three different cases: (i) without any
reconstruction of the shape, i.e. considering the
  bunch of point tracers, and approximating the particle-puff as a (ii)
  sphere or as an (iii) ellipsoid. The results show a strong
dependence on the fluctuations of the instantaneous wind at the moment
of the emission and appear to be robust with respect to the approximations (i)-(iii).
\end{abstract}

\begin{keyword}
Richardson diffusion, Lagrangian turbulence, particles dispersion.
\end{keyword}

\end{frontmatter}
\section{Introduction\label{Introduction}}

Understanding the dispersion and evolution of
  particles in turbulent flows is a fundamental problem with
  applications in different fields ranging from atmospheric and
  oceanic sciences \cite{csanady73,Bennet1984,LaMaRi2008,Ollitraut,PZ,Lacasce2010} to
  chemical engineering and astrophysics
  \cite{Baldyga,hill1976,Lepreti,Elmegreen2010} and even behavioral
  biology \cite{celani-odor_landscape}.  At high Reynolds numbers,
molecular diffusion is negligible and turbulence dominates the transport of momentum, temperature,
humidity, salinity and of all other chemical species possibly
present in the environment. Turbulent diffusion can be studied from an
Eulerian point of view, following the evolution of a concentration
field \cite{dimotakis05,Warhaft2000} or using a Lagrangian approach in
terms of particle tracers advected by the flow \cite{rev_sawford,FGV}.

In this paper we discuss an important set-up, namely when particles are emitted as a puff,
localized in time and in space. The main aim is to understand the
evolution of the shapes of these particle-puffs, quantifying the
time evolution of both the growth rate of their typical size and the
deviation from a perfect spherical shape. The evolution is followed
from the initial instant, when each particle puff is spherical and
with size $\sim\eta$, the turbulent Kolmogorov scale, up to the final time of order of one eddy turnover
time, when the bunches of particles have reached a typical scale as big as the largest eddy correlation length. We quantify the
initial distortion from the spherical shape induced by the
intermittent and intense stretching of the local gradients and the
shape's evolution for intermediate times, when the characteristic
bunch size is within the inertial range of scales. Because of the many-body nature of our experiments, one expects different informations with respect to those obtained measuring the well known two-particles
Richardson dispersion \cite{rev_sawford,rich26,rev_collins}, still anticipating however some connections between the two.

As always in passive transport problems \cite{FGV}, there is a
one-to-one dictionary that maps Lagrangian concepts into Eulerian
ones. Here, the Eulerian counterpart of puff dispersion is the
emission of a passive scalar field from a localized
source. This problem has many relevant applications, from
environmental fluid mechanics, e.g. the dispersal of noxious chemicals
in the atmosphere or in the oceans \cite{Devaull95}, to biology,
e.g. long-range olfactory communication through the emission of
volatile pheromones \cite{Wyatt03}.  The statistics and dynamics of
the concentration field at a distance from the emitting source can
thus be put in direct correspondence with puff dispersion. Notably,
the alternation of clear-air ({\it blanks}) and chemically-loaded
pockets ({\it whiffs}) that is observed away from the source is
connected with the probability that a puff hits a target at a distance
from the point where it has been delivered, and among the events where
the concentration is detectable, large intensities are due to puffs
that disperse poorly (see e.g. \cite{celani-odor_landscape} for a
detailed explanation). Turbulence, therefore, plays a major role in
determining the intensity and the temporal structure of the signal, by
shaping the information content of the odor message in the case of
olfactory communication, or by affecting the probability of reaching
lethal doses in environmental applications.

 The paper is organized as follows. In Section \ref{sec:AsphericityAndProlateness}, we introduce and analyze
the main quantities which have been used to characterize the puff
geometry, namely {\it asphericity} and {\it prolateness}, which, together with the radius of gyration, summarize the geometrical information contained in the gyration
tensor. We show that the most important signature in the two
observables is that they approach a peak value, corresponding to
maximally elongated puffs, at a time $t \sim 10 \tau_\eta$. We
understand this phenomenology as due to the physics of the dissipative
range where the puff, being still of size $\sim \mathcal{O}(\eta)$, is
controlled by the stretching rate. Indeed, we show that the initial
evolution of these observables can be reasonably estimated considering
 the bunch stretched with
exponential rates given by the three Lyapunov exponents of the
underlying fluid. Another remarkable result is obtained for longer
times, when we measure a very slow recovery of the isotropic shape, revealing that the typical puff is non spherical, at least, as
far as the evolution pertains to spatial and temporal scales in the
inertial range. In Section \ref{sec:BlanksAndWhiffs}, we exploit the link between the
Eulerian and Lagrangian language and use the statistics of puffs
hitting a given target to estimate the probability of time durations
of blanks and whiffs typical of passive scalar fields emitted from a
point source. We find that the fluctuations of the instantaneous wind
at the instant of the emission are crucial to identify the right
downstream direction where to place the target. Moreover, we show that
the probability of observing a whiff or a blank lasting for a time $t$
has a power law behavior for small times in fair agreement with the
prediction obtained in \cite{celani-odor_landscape}, using dimensional
estimate based on exit-time statistics of turbulent diffusion.

%%%%%%%%%%%%%%%%%%%%%%%%%%%%%%%%%%%%%%%%%%%%%
\section{Evolution of particle-puff geometry due to turbulence}\label{sec:AsphericityAndProlateness}

The data analyzed in this paper have been previously obtained in \cite{biferale-intermittency_in_the_relative_separations} from a 
direct numerical simulations (DNS) of the Navier-Stokes equations in
three dimensions with an isotropic and homogeneous large-scale forcing
in a tri-periodic domain. As explained in \cite{biferale-intermittency_in_the_relative_separations}, each puff is composed of $N_P$ particles,
emitted from a point-like source within the simulation box.  Particle
bunches are initially distributed uniformly in a sphere of size of
order of the Kolmogorov length scale $\eta$, and are emitted one by
one every Kolmogorov time, $\tau_{\eta}$, till about one large-scale
eddy turnover time, $T_E$, for a total of 80 emissions per source. The
total integration time of the DNS is of about $2T_E$.
Table~\ref{tab:DNS} summarizes the main parameters the DNS.
\begin{table}[phtb]\centering
\begin{tabular}{ccccccp{1\columnwidth}}
\hline
\noalign{\vskip 0.1cm}
$Re_{\lambda}$ & $N^3$ & $\eta$ & $\Delta x$ & $\varepsilon$ & $\tau_{\eta}$\\
\noalign{\vskip 0.1cm}
280&$1024^3$&0.005&0.006&0.81&0.033\\
\noalign{\vskip 0.1cm}
\hline
\noalign{\vskip 0.1cm}
$T_E/\tau_{\eta}$ & $u_{rms}$ & $N_P$ & $N_{sou}$ & $N_{puff}$ & $T_{traj}/\tau_{\eta}$\\
\noalign{\vskip 0.1cm}
80&1.7&2000&256&20480&160\\
\noalign{\vskip 0.1cm}
\hline
\end{tabular}
\caption{
Main parameters of DNS (all dimensional quantities are
  expressed in simulations units): $Re_{\lambda}$ Taylor-scale-based
  Reynolds number, $N^3$ grid resolution, $\eta$ and $\tau_{\eta}$
  Kolmogorov length and time scales, $\Delta x$ grid spacing,
  $\varepsilon$ mean energy dissipation, $\nu$ kinematic viscosity,
  $T_E$ large-scale eddy turnover time, $u_{rms}$ root-mean-square
  velocity, $N_P$ number of particle tracers within each puff,
  $N_{sou}$ number of localized sources within the simulation box, $N_{puff}$ total number of puffs in the simulation domain, $T_{traj}$ maximal duration of particle trajectories. The
  numerical domain is cubic, with periodic boundary conditions in all
  directions; a fully dealiased pseudo-spectral algorithm with second
  order Adams-Bashforth time-stepping has been used. The statistically
  isotropic and homogeneous external forcing injects energy in the
  first low-wavenumber shells, by keeping constant in time their
  spectral content, see \cite{Chen1993}.}
\label{tab:DNS}
\end{table}

\begin{figure}[!hbt]
\centering 
\includegraphics[scale=0.7]{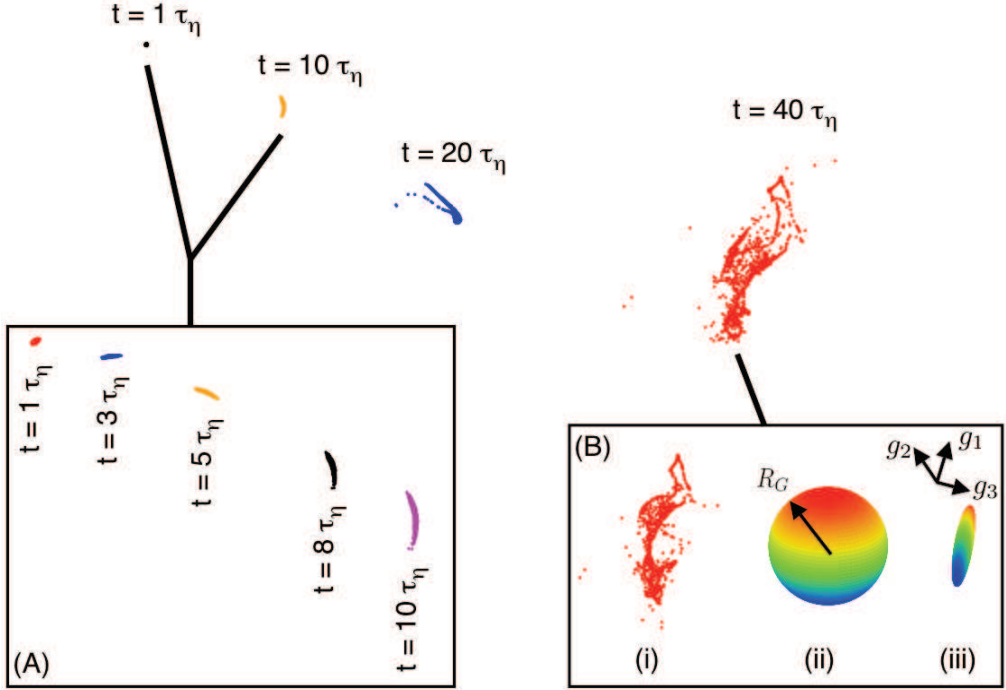}
\caption{Evolution of a given puff in space, showed at four different instants of time between $t=1\tau_\eta$ and $t=40\tau_\eta$. (A): the same puff evolution showed between $t=1\tau_\eta$ and $t=10\tau_\eta$, to illustrate how the puff undergoes a big deformation at small times, passing from a spherical shape to an elongated one. (B): the puff for $t=40\tau_\eta$ in the main picture is here represented (i) with the
  individual particles composing it, (ii) approximated with a sphere of
  radius $R_G$ (the black arrow) or (iii) approximated with an
  ellipsoid of axes $\sqrt{g_i}$, with $i=1,2,3$,  and orientation given by the direction of the three eigenvectors of the gyration tensor (the three black arrows). Both the sphere and the ellipsoid are centered in the center of mass of the puff.}
\label{fig:approximation}
\end{figure}

Once emitted, the particle puffs are transported and deformed by the
turbulent flow. At least at low-order, we can describe their main
geometrical features by measuring the gyration tensor,
$\textbf{G}(t)$, with elements
\begin{equation}
G^{ij}(t)=\frac{1}{N_P}\sum_{n=1}^{N_P} (r^i_n(t)-r^i_{CM}(t))(r^j_n(t)-r^j_{CM}(t)), \ \ \ i,j=x,y,z
\label{gyration tensor}
\end{equation}
where $r^i_n(t)$ is the $i$-th coordinate of the $n$-th particle of a
given bunch after a time $t$ from its emission, and the subscript
${\it CM}$ stands for the center of mass. In particular, the three
eigenvalues, $\{g_{i}(t)\}_{i=1}^{3}$, of $\textbf{G}(t)$ bear
information about both the size and the shape of the puffs (see, e.g., \cite{Cher1999, toschi-lagrangian_properties}). For instance,
their sum, corresponding to the trace of the gyration tensor, provides the squared radius of gyration $R_G^2(t)$, namely the bunch
characteristic size approximated with a sphere of radius $R_G(t)$:
\begin{equation}
\sum_{i=1}^3 G^{ii}(t)=\sum_{i=1}^{3} g_i(t)=R_G^2(t)\
\label{eigenvalues}
\end{equation}
On the other hand, the three eigenvalues corresponds to the square of
the semi-axes of the ellipsoid which best approximates the particle
bunch. In Figure~\ref{fig:approximation} we show a schematic picture of the evolution of a given puff, and we explain the link between the bunch shape and both $R_G$ and the eigenvalues $g_i$.

\begin{figure}[!hbt]
\centering 
\includegraphics[scale=0.45]{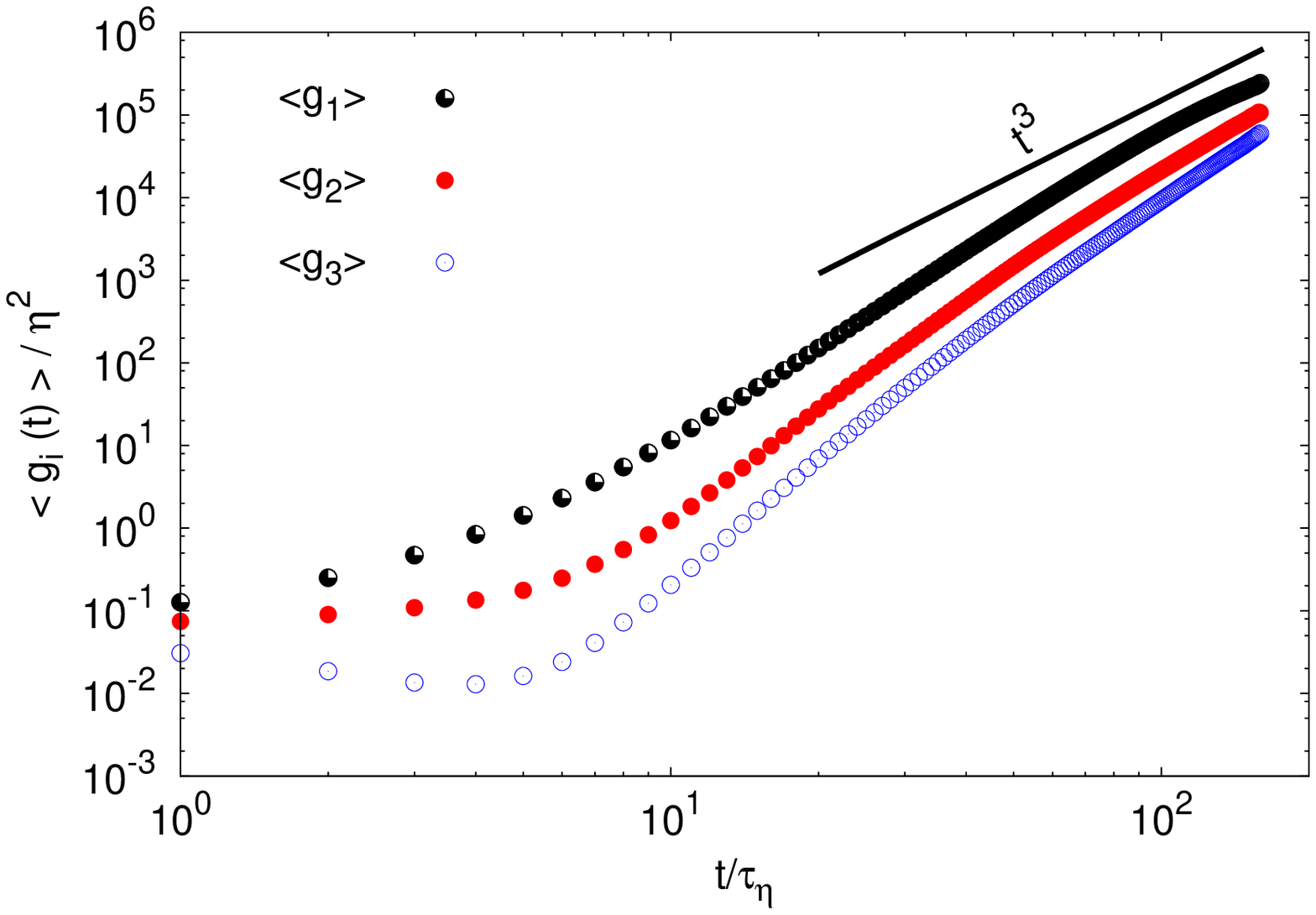}
\caption{Log-log plot of the evolution of the mean eigenvalues $\langle g_1\rangle$, $\langle g_2\rangle$ and $\langle g_3\rangle$ of the gyration tensor
  $\textbf{G}(t)$ versus time. The straight line is proportional to
  $t^3$ and corresponds to the Richardson law (\ref{eq:rich}). The three eigenvalues start with very close values (not
  shown), typical of a spherical bunch, and then separate quickly, to
  end up with slightly different values, which means that on average
  the bunch does not return to a spherical
  shape.}
\label{eigen_evolution}
\end{figure}

A first glimpse into the evolution of puffs  can be obtained by looking at the
behavior of the three eigenvalues averaged over all sources and
emissions, $\langle g_i(t) \rangle$, as a function of the time from
their emission (Figure~\ref{eigen_evolution}).
Two regimes are easily
detectable. At short times, when the puff size is below or comparable to the
Kolmogorov length scale, the dynamics is expected to be controlled by
the dissipative range physics and thus by the exponential stretching
rates. In other terms, ignoring effects due to stretching rates
fluctuations \cite{fujisaka83,benzi85}, we can approximatively assume
that the eigenvalues $g_i$ are linked to the Lyapunov exponents
$\lambda_i$ through the relation $\lambda_i\approx
\ln\langle g_i(t)\rangle /2t$. The factor 2 in the denominator
comes from the fact that the gyration tensor is quadratic in particle
separations. The Lyapunov exponents characterizing the dynamics of
particle tracers in turbulent flows have been measured in several
studies \cite{pope90,Bif2005b,BecPoF2006}, here we refer to the values
measured in \cite{BecPoF2006}: $\lambda_1 \sim
0.14/\tau_\eta$, $\lambda_2 \sim 0.04/\tau_\eta$, and $\lambda_3=
-\lambda_2-\lambda_1$, as $\Sigma_i\lambda_i=0$, due to fluid
incompressibility. In Figure \ref{eigen_evolution} we can see that
$\langle g_1(t)\rangle$ grows faster than the other two, being
associated with $\lambda_1$, $\langle g_2(t)\rangle$ grows slower,
whereas $\langle g_3(t)\rangle$ decreases initially due to the fact
that $\lambda_3$ is negative, indicating that at the early stage of
the evolution the spherical bunch tends to preserve the volume while
fluid gradients deform it.

At a later time, for $t \gtrsim 10 \tau_\eta$ all three typical sizes
are inside the inertial range, we would then expect the Richardson
scaling for all eigenvalues \cite{Bif2005b,Pumir2000}:
\begin{equation}
\langle g_i(t) \rangle= g_R \epsilon t^3
\label{eq:rich}
\end{equation}
where $g_R$ is the {\it universal} Richardson constant and $\epsilon$
is the mean energy dissipation of the fluid.  As one can see from the
figure, this scaling behavior is barely achieved only for large times
in agreement also with the behavior of the eigenvalues of the inertia
tensor for particle tetrads~\cite{Bif2005b}. This deviation from
Richardson can be explained as due to the contamination
induced by the fluctuations of the viscous scale leading to the
presence of bunches of various sizes (even of order $\eta$) also at long times after the emission, see
\cite{biferale-intermittency_in_the_relative_separations,SBT12} for a
detailed study of this effects on the same data set. It is interesting
to remark, that the three curves tend to proceed parallel when inside
the inertial range. They show a tendency to converge to the same value
only at very long times when the dynamics is essentially diffusive, as
above the largest scale of turbulence velocity correlations are washed
out.  This means that in the inertial range the single puff,
either never recovers isotropy (i.e. a spherical shape) or it does it very
slowly. This is an important observation as it tells us that it would
be wrong to describe a typical puff evolving in a turbulent flow in terms of 
a spherical shape, even if the underlying turbulence is isotropic. 
 
In the remainder of this section we quantify these deviations from
spherical symmetry. To this aim we adopt two well known 
quantities~\cite{aronovitz-universal_features,blavatska-shape_anisotropy} defined in terms of the eigenvalues of $\textbf{G}(t)$, the 
{\it asphericity} and {\it prolateness}.  The {\it
  asphericity} is given by the normalized variance of the three
eigenvalues,
\begin{equation}
A(t)=\frac{1}{6}\sum_{i=1}^3\frac{(g_i(t)-\overline{g(t)})^2}{\overline{g(t)}^2}
\label{asphericity}
\end{equation}
where $\overline{g(t)}= \sum_{i=1}^{3} g_i(t)/3$.  Clearly, we have that $A$ ranges from $0$, when
all eigenvalues are equal and the bunch is spherical, to 1 with only
one eigenvalue different from zero, meaning a totally aspherical
bunch. The {\it prolateness}:
\begin{equation}
P(t)=\frac{\prod_{i=1}^3(g_i(t)-\overline{g(t)})}{\overline{g(t)}^3}\
\label{prolateness}
\end{equation}
is defined in the interval $[-1/4,2]$ and tells us whether the shape
of the bunch is oblate or prolate. In particular, $P(t)>0$ means
$g_1\gg g_2\approx g_3$, i.e. the bunch has a prolate shape becoming a rod in
the limit $P=2$, while $P(t)<0$ means $g_1\approx g_2\gg g_3$ that is
the bunch is oblate, tending to a disk for $P=-1/4$.

\begin{figure}[!hbt]
\centering 
\includegraphics[scale=.55]{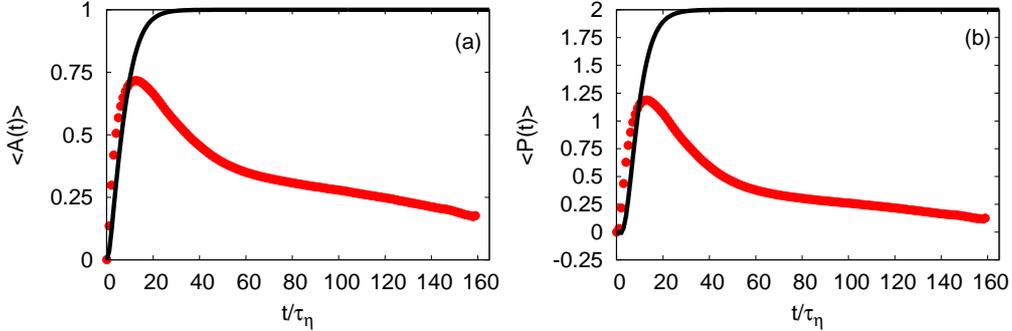}
\caption{(a): Lin-Lin plot of the mean asphericity $\langle
  A(t)\rangle$ versus time (points). The value at time $t=0$ is very small $\approx 10^{-3}$, because the bunch is injected with a
  spherical shape.  At the end of its evolution $\langle A(t)\rangle$
  does not return close to zero, that is the bunch does not recover a
  spherical shape. The solid black line approximating the initial growth
  is evaluated assuming the Lyapunov dynamics for the eigenvalues,
  i.e. replacing $g_i(t)$ with $e^{2\lambda_it}$ (see text for the
  Lyapunov values). The peak in the real evolution of the
  asphericity corresponds to $ t/\tau_\eta \approx 13$, i.e. the time when the size of the puffs starts to be dominated by
  non-linear effects. (b): the same discussion for the asphericity applies to the prolateness.}
\label{fig:asphericity}
\end{figure}
In Figure~\ref{fig:asphericity}a-b we show the evolution of the
asphericity and prolateness, averaged over all emissions, $\langle
A(t) \rangle$ and $\langle P(t) \rangle$, respectively.  As at the
emission time $t=0$ each particle bunch is injected uniformly within a
sphere of radius $\sim \mathcal{O}(\eta)$, the initial values are close to zero
$\langle A(0)\rangle\sim \mathcal{O}(10^{-3})$ and $\langle
P(0)\rangle\sim \mathcal{O}(10^{-6})$.

At initial times, while the bunch size is within the dissipation range
(for $t \lesssim 10 \tau_\eta$), the asphericity grows very rapidly,
reaching a peak value $\approx 0.7$ at $t\approx 13\tau_\eta$. This
means that the dissipative scale stretching mechanism is very
efficient in transforming the sphere in a prolate ellipsoid, as
confirmed by the average prolateness (Fig.~\ref{fig:asphericity}b)
which reaches a peak value $\approx 1.2$ when $\langle A(t)\rangle$ is
maximal. This kind of evolution is indeed expected due to the presence
of two positive Lyapunov exponents \cite{pope90,Bif2005b,BecPoF2006}.
As shown in Figure~\ref{fig:asphericity}, the initial growth of both asphericity and prolateness
is well captured by their approximations obtained from
(\ref{asphericity}) and (\ref{prolateness}) by replacing $g_i(t)$ with
$e^{2\lambda_it}$. We notice that the actual data grow slightly faster
than that predicted by the Lyapunov exponent approximation, this
discrepancy might be due to the fact that we neglected fluctuations of the
stretching rates \cite{fujisaka83,benzi85}. Notice that at time long
enough the (linear) approximation reaches the asymptotic values
typical of a rod like structure, as the approximation does not
include neither folding mechanism nor the physics of the inertial
range with the Richardson diffusion.

After the peak, both the asphericity and the prolateness show a slow
decay toward lower values, but even for times larger than the large-scale eddy turnover time, the shape of the bunch does not return spherical. Clearly, at very long times, where only diffusion is
taking place, we expect to recover a spherically symmetric puff, but
this does not seem to be the case within the inertial range.

\begin{figure}[!hbt]
\centering
\includegraphics[scale=0.55,angle=0]{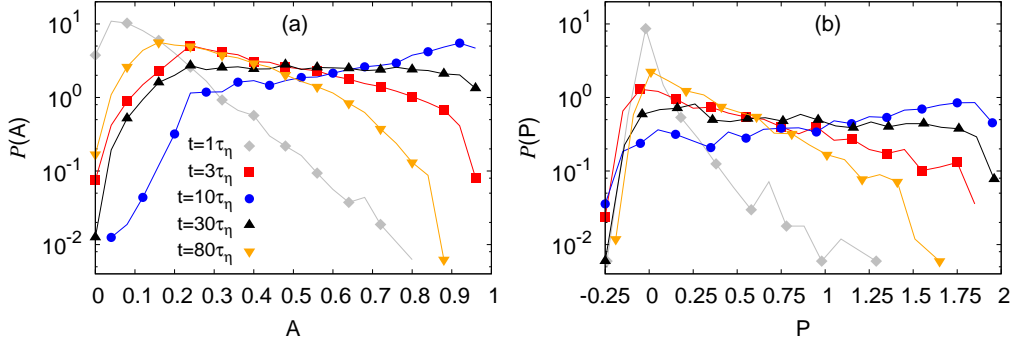}
\caption{Probability density functions for the asphericity (a) and the prolateness (b), at various times, $t=1\tau_\eta (\Diamond)$, $t=3\tau_\eta (\square)$, $t=10\tau_\eta (\bigcirc)$, $t=30\tau_\eta (\bigtriangleup)$, $t=80\tau_\eta (\bigtriangledown)$. Even at large times there is still a considerable number of puffs far from having a spherical shape.}
\label{a_p_pdf}
\end{figure}
In Figure~\ref{a_p_pdf}a-b we show the probability density functions (PDFs) for the asphericity
and the prolateness, respectively, at various times.  After a time
interval of approximately $ t \sim 10 \tau_\eta$, that corresponds to
the peak values of Figure \ref{fig:asphericity}a-b, we observe that turbulent stretching has produced
bunches with all possible values of $A$ and $P$ with a higher
probability for very elongated ellipsoid. For later times, a slow
return toward the spherical values is observed, with the development
of a peak at $A =P=0$. Notice however that, even for very large times,
both PDFs show the presence of a large sub-set of bunches that are far
from being spherical, a clear and important legacy of small-scale
turbulent fluctuations for the whole time history of every bunch.

%%%%%%%%%%%%%%%%%%%%%%%%%%%%%%%%%%%%%%%%%%%%%%%%
\section{Blanks and whiffs \label{sec:BlanksAndWhiffs}}

 In view of the presence of a non negligible set of
  bunches with a non-spherical shape one might want to assess the
  effects of shape distribution on the probability of hitting a target
  and how this impacts the concentration statistics in a given point
  downstream of the emitting source.  In
  particular, we are interested to compare three different cases: (i)
  considering the puff as formed by individual particles, (ii)
  approximating it as a sphere with a radius given by its mean radius
  of gyration or (iii) approximating it as an ellipsoid with the three
  main axes given at any time by the value of the three eigenvalues
  $g_i(t)$.  Of course the probability to hit the target will strongly
  depend on the presence (or absence) of a mean wind, on the distance
  between the source and the target and on the wiggling motion of the
  center of mass of the puff induced by turbulent fluctuations. In the
  present case, because of the absence of a steady mean wind, we have
  always conditioned the position downstream of the target to be
  aligned with the direction of the velocity in the position of the
  source at the instant of the emission. We also adopted a definition of (local)
  mean wind based on the average displacement of the center of mass of all 
  puffs emitted by the same source. We found that the results do not depend 
  too strongly on the adopted definition provided the target is at distance lower 
  than the large scale of the flow. Indeed the local wind being a large scale quantity
   will persist on the time scale of the largest length scale. In absence of this fine
  tuning, the probability to hit the target decreases to values that
  are not statistically significant (not shown). For sake of definiteness, the
  target has always been taken spherical, of radius $R=4\eta$ and
  placed at distance $D=40\eta$ from the source, along the direction
  given by the (local) mean wind.
\begin{figure}[!hbt]
\centering
\includegraphics[scale=0.4]{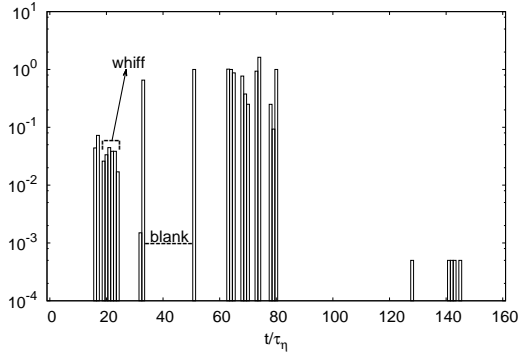}
\caption{Histogram of the signal of the concentration of particles for
  a single source. The target is placed at a distance $D=40\eta$
  from the source, along the direction given by the mean wind, and all the
  values of concentration are divided by the total number of particles
  that compose a puff. Periods of presence of signal (\textit{whiffs}) and of absence of signal (\textit{blanks}) can be distinguished.}
\label{signal}
\end{figure}
Notwithstanding the huge data set, the lack of a stable mean wind
makes this kind of measurement very delicate, as the puff is advected
by the mean velocity of its center of mass, which is a highly
fluctuating quantity if the puff has a relatively small size. As a
result only rarely the bunches hit the target and the signal given by
the superposition of the puff with the target is very
intermittent. The only stable probability we could measure is the time
elapsed between two consecutive hits (the time span of a blank $t_b$), and
the time duration of a detection (the time span of a whiff $t_w$), 
independently of the amount of particles detected. Both times are
predicted to have a power law distribution at least for short
durations, when the determination of a blank or a whiff is due to the
three-dimensional diffusive motion of the center of mass of the bunch
around the target \cite{celani-odor_landscape}:
\begin{equation}
\label{eq:pdfwhiff}
P(t_b) \sim t_b^{-3/2}; \qquad P(t_w) \sim t_w^{-3/2}
\end{equation}
 According to this picture, the beginning of a whiff occurs when the
 puff is barely within the target, i.e. the center of the bunch is at
 a distance of order $R_G$ from the target. The end of the whiff will
 occur when the puff first exits from the target.  Therefore, the
 duration is distributed as the first exit time for a diffusion
 process, whence the power law $3/2$ in (\ref{eq:pdfwhiff}). For
 events of a longer duration an exponential tail should emerge (see
 \cite{celani-odor_landscape} for a detailed modelization at all
 times). Similarly, the duration $t_b$ of blanks is the time needed
 for a diffusing puff that has lost contact from the target to regain
 contact with it.  It follows from arguments symmetric to those
 discussed for the whiffs that blank intervals are distributed as
 $t^{-3/2}$ as well.
 
 The signal of the concentration of particles
 within the target for a specific source is presented in Figure
 \ref{signal}. The heights of the histogram represent the values of
 concentration divided by the total number of particles in a
 bunch. These are evaluated integrating over all the time the bunch
 remains within the target. The signal is intermittent and very
 similar to those obtained experimentally, and we can distinguish
 periods of absence of signal (blanks) and periods of presence of
 signal (whiffs). Blanks correspond to the case in which the bunch
 does not intersect the target and passes by. On the contrary whiffs
 correspond to the case in which the bunch is partially or entirely
 inside the target. As from equation (\ref{eigenvalues}) we can
 consider the bunch not only as a discrete object composed of
 particles, but also as a sphere of radius $R_G$ or as an ellipsoid
 with semi-axes given by $\sqrt{g_1}$, $\sqrt{g_2}$ and
 $\sqrt{g_3}$. In all these three cases, we can reconstruct the
 probability density functions of blanks and whiffs.
\begin{figure}[!hbt]
\centering
\includegraphics[scale=0.55]{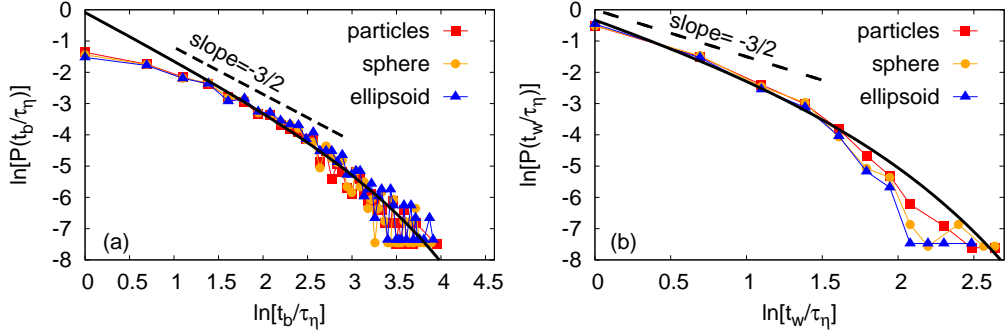}
\caption{PDF of the duration $t_b$ of blanks (a), i.e. time intervals without
  signal, and PDF of the duration $t_w$ of whiffs (b), i.e. time intervals with signal, for the bunch considered as composed of particles
  ($\square$) and approximated with a sphere ($\bigcirc$) and an
  ellipsoid ($\bigtriangleup$). On the axes, the natural logarithms of $t_b$ $(t_w)$ and $P(t_b)$ $(P(t_w))$ are reported. The dashed line has slope equal to -3/2 and represents the expected trend for small times, $P(t_b)\sim t_b^{-3/2}$ for blanks and $P(t_w)\sim t_w^{-3/2}$ for whiffs, while the solid line represents the expected
  trend for the entire PDF, $P(t_b)\sim t_b^{-3/2}e^{-t_b/T}$ for blanks and $P(t_w)\sim t_w^{-3/2}e^{-t_w/T}$ for whiffs. The PDFs are obtained considering a threshold value for concentration, i.e. not considering in the signals, as for instance the one shown in Figure \ref{signal}, heights smaller than $10^{-2}$.}
\label{blank_whiff}
\end{figure}
The PDF of the blanks is shown in Figure \ref{blank_whiff}a, for the bunch
taken as discrete and approximated with a sphere and an ellipsoid. The dashed 
black line represents the law we expect the three cases
will follow for small $t_b$s, i.e. (\ref{eq:pdfwhiff}), while the solid black line is the expected trend for the entire PDF, i.e. equation (\ref{eq:pdfwhiff}) times an exponential term $\sim e^{-t_b/T}$. 
Very short blanks tend to be less frequent, as it is unlikely that a puff crosses the target, leaves it and then hits back 
after a time of order $\tau_\eta$. The PDF of the whiffs is instead shown in Figure \ref{blank_whiff}b. Since blanks and whiffs are complementary, we expect the same law. In this case we expect instead few whiffs of long duration. Even without a well defined mean wind we could have obtained satisfying results, which suggest blanks and whiffs are two
robust observables. As far as our statistics is concerned, using the
real discrete shape, the spherical or ellipsoidal approximation has no
consequences regarding the probability distribution functions. If
this remains the case also with a larger statistics and/or in presence
of a mean stable wind is an interesting and open question that we
cannot answer within the present data set. However, the argument used 
in \cite{celani-odor_landscape} (and briefly discussed above) to derive (\ref{eq:pdfwhiff}) 
is rather general depending on the diffusive motion of the center of mass and thus the 
behavior (\ref{eq:pdfwhiff}) is not expected to depend too strongly on the representation 
of the puff.

%%%%%%%%%%%%%%%%%%%%%%%%%%%%

\section{Conclusions}

We have studied the evolution of shape of more than $10^4$ different puffs transported by a turbulent flow, by means of asphericity and prolateness, following the puffs up to one large-scale eddy turnover time. The puff is emitted spherical in the flow, and we have found a strong distortion of the shape at small times, that is also in agreement with assuming the Lyapunov dynamics for the characteristic dimensions of the puff. Moreover, we have found that notwithstanding the isotropic nature of turbulence, puffs do not fully recover a spherical shape at longer times for a threefold reason:
(i) small-scale stretching quickly and strongly distorts the puff (ii) the recovery of 
isotropy in the inertial range is slow (algebraic) \cite{aniso} (iii) the inertial range has a limited extent.
As a result, the memory of the Lagrangian dynamics in the dissipative range is kept 
for long times. We have also analyzed the probability of hitting a given target placed downstream with respect to the local wind at the time of emission, in the three cases of the puff considered as a discrete bunch of particles, as a sphere, and as an ellipsoid. We have found a good agreement with the predictions for the probability densities of time periods with and without hits. These results point to the conclusion that the concentration statistics is only mildly affected by the asphericity of the puffs. Our analysis, possibly extended to include the effect of mean wind and shear, 
may provide valuable suggestions to improve Lagrangian models of puff release in the 
environment \cite{haan1995,haan1998,reynolds2000}.

%%%%%%%%%%%%%%%%%%%%%%%%%%%%%%%%

\section*{Acknowledgements}
L.B. acknowledges the collaboration with R. Scatamacchia, A.S. Lanotte and F. Toschi for the production of the data analyzed in this work and the 
partial funding from the European Research Council under the European Community's Seventh Framework Programme, ERC Grant Agreement N. 339032.

%%%%%%%%%%%%%%%%%%%%%%%%%%%%%%%

\section*{References}

\bibliographystyle{jfm}

\end{document}